\documentclass[aps,showpacs,amsmath,amssymb,twocolumn,pra]{revtex4-1}
\usepackage{bm}
\usepackage[dvips]{graphicx}
\usepackage{wrapfig,subfigure}
\usepackage{amssymb}
\usepackage{color}
\usepackage[normalem]{ulem}
\usepackage[caption=false]{subfig}
\usepackage{amsmath}
\usepackage{epstopdf}

\begin{document}
\title{Anomalous dissipation of nanomechanical modes going through nonlinear resonance} 
\author{O. Shoshani}
\affiliation{Ben-Gurion University of the Negev, Beer-Sheva 8410501, Israel}
\author{S. W. Shaw}
\affiliation{Florida Institute of Technology, Melbourne, FL 32901, USA}
\author{M. I. Dykman}
\affiliation{Michigan State University, East Lansing, MI 48823, USA}

\date{\today}

\begin{abstract}
We study nonlinear resonance of coupled modes in nano-mechanical systems. To reveal the qualitative features of the dynamics, we consider the limiting cases, where the results can be obtained analytically. For 1:3 resonance, we find the anomalously strong and nonmonotonic dependence of the decay rate of the low frequency mode on its amplitude, if the decay rate of the high-frequency mode is comparatively large. In this case the low-frequency mode driven close to resonance can have several branches of steady-state vibrations with constant amplitude. If the decay rates of the both modes are small compared to their coupling and internal nonlinearity, the dynamics corresponds  to slowly decaying strongly nonsinusoidal oscillations of the vibration amplitude. Weak driving can make these vibrations stable.
\end{abstract}
\maketitle

\section{Introduction}
Nonlinear resonance is a problem with long history in quantum and classical mechanics, going back to Laplace and Poincare on the classical side and to the Fermi resonance on the quantum side \cite{Arnold1989,Fermi1931}. It occurs in a broad range of systems, from celestial bodies to ecological systems to molecules \cite{nayfeh1973nonlinear,alfriend1971stability,blasius1999complex,Nakamoto2008}. Recently nonlinear resonance has attracted particular interest in the context of nano-and micromechanical vibrational systems \cite{Eichler2012,Antonio2012,Matheny2013,Mahboob2013,qalandar2014frequency,nitzan2015self,Mangussi2016,Guettinger2016,Chen2016} as well as the microwave cavities used in quantum information. These systems provide an unprecedented access not just to studying this complicated phenomenon, but also to controlling and using it.

In conservative classical systems, nonlinear resonance leads to the onset of nonlinear vibrations accompanied by energy oscillations between the resonating modes. This is reminiscent of linear resonance between two harmonic oscillators.  However, on a finer scale, the picture is more complicated, extending to dynamical chaos. On the quantum side, nonlinear resonance is in some sense simpler in the absence of dissipation, as it leads to the standard level repulsion. 

The quantum situation changes if the resonating modes are dissipative. If the modes have very different decay rates, one of them can serve as a thermal reservoir for another \cite{Dykman1978}. This effect has attracted much attention in cavity optomechanics \cite{Aspelmeyer2014a} and has been recently used to drive a slowly decaying microwave cavity mode to a coherent quantum state \cite{Leghtas2015}. However, this behavior can change dramatically when the mode nonlinearity is significant \cite{Sun2016}.

An important advantageous feature of mesoscopic oscillators is the possibility to tune the modes in and out of nonlinear resonance. This can be done statically, by directly controlling the mode frequencies \cite{Eichler2012}, and  also dynamically, by taking advantage of  the dependence of the mode eigenfrequency on the vibration amplitude. By driving a mode close to its eigenfrequency, and thus increasing its vibration amplitude, one can bring its overtone in resonance with an overtone of another mode. This second mode is then excited, too. The backaction from this excitation can significantly change the dynamical response of the driven mode. 

In this paper we develop a theory of weakly damped vibrational modes brought into nonlinear resonance. A specific feature of the nonlinear resonance is that both the strength of the nonlinear coupling and the mode frequencies depend on the modal amplitudes. Therefore a fast decaying mode can be an efficient thermal reservoir for another mode in one range of the amplitudes, but not in the other. One can then expect a strongly non-exponential decay of the slower-decaying mode. On the other hand, if in a certain amplitude range the decay is weaker than the coupling in the appropriate units,  the dynamics corresponds to resonant inter-mode energy oscillations. However, as the amplitudes decay, the scaled coupling can become weaker than the decay, eliminating the oscillations.

The damping of mesoscopic modes is usually weak, the decay rate is small compared to the mode frequencies. Such modes can be efficiently excited  by a comparatively weak driving if it is resonant. This is advantageous, as it allows keeping the leading terms responsible for the resonant behavior while eliminating irrelevant variables. Respectively, we will describe the dynamics of the coupled modes in the rotating wave approximation (RWA).

Our study aims at addressing two problems: the decay of modes that go through nonlinear resonance and the response of the modes to a resonant field. We will concentrate on the limiting cases. The general analysis can be only done numerically. Extensive numerical simulations have been done in parallel with this paper, along with the experimental studies simultaneously and independently by two groups \cite{Guettinger2016,Chen2016}.

To be specific, we will consider the modes with frequency ratio close to 1:3. In many nano- and micromechanical systems the coupling between such modes is comparatively strong; for symmetry reasons, it is often stronger than the coupling between the modes close to 1:2 resonance. The onset of 1:3 nonlinear resonance in a micromechanical system, which is similar to the one we use here, was inferred previously from indirect data, but the related nonlinear dynamics was not explored \cite{Antonio2012}.  The work \cite{Guettinger2016,Chen2016} also refers to 1:3 resonance.


One of the specific limiting cases we will address are the case where one of the modes has a much higher decay rate than the other. In this case the dynamics can be described in the adiabatic approximation, in which the fast decaying mode follows the evolution of the slowly decaying mode adiabatically. This leads to an anomalous nonlinear friction for the slowly decaying mode. Along with the nonlinear friction there arises a change of the effective vibration frequency of the slowly decaying mode, that displays a pronounced and strongly  nonmonotonic dependence on the vibration  amplitude. This is the adiabatic analog of the nonlinear frequency anticrossing. 

The other specific case is the case where the decay rates of the both modes are small compared to the appropriately scaled nonlinear mode coupling and possibly to the nonlinear shift of the mode frequencies compared to the small-amplitude eigenfrequency values. In this case the dynamics in the absence of the drive and dissipation is fully integrable, in the RWA. In the rotating frame, it maps onto vibrations of a ``pendulum." We will describe the effect of weak dissipation and a weak field, which in this case can lead to the onset of limit cycles in the rotating frame, with the amplitude determined by the competition between energy drain due to the dissipation and the energy gain from  the field.

\section{The rotating wave approximation for coupled modes}

To explore the driving-tuned nonlinear resonance it suffices to study two coupled modes. The modes can be described by the simplest model that displays the amplitude dependence of the vibration frequency, the Duffing oscillator. In the absence of coupling to the environment that leads to dissipation and fluctuations, the Hamiltonian of the modes reads $H = H_1 + H_2 + H_{\rm cpl} + H_{F}$ with
\begin{align}
\label{eq:Hamiltonian}
& H_1 = \frac{1}{2}p_1^2 + \frac{1}{2}\omega_{1}^2q_1^2 + \frac{1}{4}\gamma_{11} q_1^4,\nonumber\\
&H_2 = \frac{1}{2}p_2^2+ \frac{1}{2}\omega_{2}^2q_2^2 +\frac{1}{4}\gamma_{22}q_2^4, \nonumber\\
&H_{\rm cpl} = \gamma_{12}q_1^3q_2, \quad H_F = -q_1 F\cos(\omega_Ft),
\end{align}
Here, $q_{1,2}$ and $p_{1,2}$ are the coordinates and momenta of the modes, $\omega_{1,2}$ are the mode eigenfrequencies, $\gamma_{11}$ and $\gamma_{22}$ are the parameters of the Duffing nonlinearity, $F$ and $\omega_F$ are the amplitude and frequency of the driving field, and  $\gamma_{12}$ is the mode coupling parameter.  The model (\ref{eq:Hamiltonian}) captures the essential features of the resonance.  In the considered regime, other nonlinear terms renormalize the parameters. 

We assume that $\omega_F$ is close to the mode-1 eigenfrequency, $|\omega_F -\omega_1| \ll \omega_F$. We also assume that $3\omega_1$ is close to $\omega_2$. However, the detuning $|3\omega_1 - \omega_2|$, even though it is small compared to $\omega_1$, is not necessarily small compared to the coupling strength, so that one needs the driving to tune modes in resonance. 

The driving-induced tuning occurs because the frequency of the mode 1 depends on its amplitude,  $\omega_{\rm 1 eff} \approx \omega_1 + (3\gamma_{11}/4\omega_1)\overline{q_1^2}$, where the bar means averaging over the vibration period. If $(\omega_2-3\omega_1)/\gamma_{11} > 0$ then, by increasing $\overline{q_1^2}$, one can make $3\omega_{\rm 1 eff}$ coincide with $\omega_2$. For concreteness we set $\gamma_{11},\omega_2-3\omega_1 >0$; respectively, of interest is the range where the frequency detuning of the field 
$\omega_F-\omega_1 >0$, so that the frequency of the forced vibrations of mode 1 could resonate with $\omega_2/3$.

To analyze the effect of resonant driving, we can use the rotating wave approximation (RWA) and change from the coordinates and momenta of the modes to new scaled coordinates and momenta, $q_j = \omega_j^{-1/2} (Q_j\cos \phi_j + P_j\sin \phi_j), p_j = -\omega_j^{1/2}(Q_j\sin \phi_j- P_j\cos \phi_j)$ with the phases $\phi_1 = \omega_Ft$ and $\phi_2=3\omega_Ft$. We use subscript $j=1,2$ to enumerate the dynamical variables  of the modes 1, 2. Functions $Q_j,P_j$  are slowly varying in time, they remain almost constant over the driving period $2\pi/\omega_F$. The equations of motion for these functions are
\begin{align}
\label{eq:RWA}
&\dot Q_j= -\Gamma_jQ_j+\frac{\partial H_{\rm RWA}}{\partial P_j}, \qquad \dot P_j =-\Gamma_jP_j -
\frac{\partial H_{\rm RWA}}{\partial Q_j} \nonumber\\
&H_{\rm RWA} = -\sum_j \left[\frac{1}{2}\delta\omega_j (Q_j^2 + P_j^2) -\frac{3\gamma_{jj}}{32\omega_j^2}(Q_j^2+P_j^2)^2 \right]
\nonumber\\
&-\frac{F}{2\sqrt{\omega_F}}Q_1+
\frac{\gamma_{12}}{8\sqrt{3}\omega_F^2}{\rm Re}\,[(Q_1-iP_1)^3(Q_2+iP_2)]
\end{align}
Here, we have incorporated the terms proportional to $\Gamma_1$ and $\Gamma_2$ that describe the decay of the modes. In the phenomenological model of the mode dynamics they correspond to the friction forces $\Gamma_j\dot q_j$ that drive the modes; incorporating them into Eq.~(\ref{eq:RWA}) is justified even where the friction is delayed in the lab frame. The frequency detunings are
\[\delta\omega_1 = \omega_F - \omega_1,\qquad \delta\omega_2 = 3\omega_F -\omega_2,\]
and we have disregarded corrections $\sim |\delta\omega_{1,2}|/\omega_F$.

A major feature of the driving-induced nonlinear resonance seen from Eq.~(\ref{eq:RWA}) is that it can be described by autonomous equations of motion. They have a Hamiltonian part, with the Hamiltonian $H_{\rm RWA}$, and  a part that describes dissipation. The coupling between the modes is described by the last term in $H_{\rm RWA}$. 

\section{Resonant nonlinear friction:\\ The adiabatic regime}
\label{sec:nonlinear_friction}

The dynamics of coupled modes becomes particularly simple if mode 2 decays much faster than mode 1, $\Gamma_2\gg \Gamma_1$. Physically, mode 2 can serve as a thermal reservoir for mode 1. This ``reservoir" has two important features. First, it has a finite bandwidth. In the considered 1:3 resonance, the reservoir is efficient if the vibration frequency of mode 1 is close to $\omega_2/3$. Second, the coupling to the reservoir is nonlinear in the mode-1 coordinate. Therefore one may expect that the resulting dissipation rate will depend on the amplitude of mode 1. 

If the appropriately weighted mode coupling is small compared to $\Gamma_2$ and $\Gamma_2\gg \Gamma_1$, mode 2 follows mode 1 adiabatically. The restriction on the coupling is the analog of the familiar condition of a weak coupling to a thermal reservoir. Because of that, the vibration amplitude of mode 2 is comparatively small. One can then disregard the internal Duffing nonlinearity of mode 2, i.e., to set $\gamma_{22}$ equal to zero in Eq.~(\ref{eq:RWA}). 

To gain insight into the ensuing nonlinear friction of mode 1 we will consider first the dynamics in the absence of driving, $F=0$. The problem here is how mode 1 decays from the initial value of its amplitude in the presence of resonant nonlinear coupling. 

It is convenient to introduce a dimensionless complex amplitude of mode 1,
\begin{align}
\label{eq:amplitudes}
&A_{\rm ad}(t) = \left(\frac{9\gamma_{11}}{8\omega_1^2\Gamma_2}\right)^{1/2}[Q_1(t)-iP_1(t)]e^{-i\Phi_{\rm ad}(t)},\nonumber\\
&\Phi_{\rm ad}(t) = (\Gamma_2/3)\int^t dt' |A_{\rm ad}(t')|^2 - \delta\omega_1 t.
\end{align}
In the adiabatic regime, one can solve the equation of motion for mode 2 by setting $Q_2-iP_2 = A_2\exp[3i\Phi_{\rm ad}(t)]$ and disregarding $\dot A_2$ compared to $\Gamma_2A_2$. This gives a simple linear algebraic equation for $A_2$. Using its solution, we obtain from Eq.~(\ref{eq:RWA})
\begin{align}
\label{eq:adiabatic}
&\dot A_{\rm ad} = -\Gamma_1 A_{\rm ad}{\cal R}(|A_{\rm ad}|^2), \quad 
{\cal R}(x)=1 + \frac{\zeta_{\rm ad} x^2}{1 + i(\delta\Omega_{12} +x)},\nonumber\\
&\zeta_{\rm ad} = (\gamma_{12}/9\gamma_{11})^2\frac{\Gamma_2}{\Gamma_1},\qquad \delta\Omega_{12} = \frac{3\omega_1 - \omega_2}{\Gamma_2}.
\end{align}
This expression applies for $|\dot A_{\rm ad}|\ll \Gamma_2 |A_{\rm ad}|$, a condition that imposes a limitation on the vibration amplitude and the coupling strength. For the period-averaged displacement $\overline{q_1^2}$ of mode 1 it reads  $(\gamma_{12}^2/9|\gamma_{11}|)\overline{q_1^2} \ll 4\omega_1\Gamma_2$. 

Equation (\ref{eq:adiabatic}) shows that the backaction from the fast-decaying mode 2 leads to a nonexponential decay of the vibration amplitude of mode 1. The decay rate is maximal where the amplitude-dependent frequency of mode 1, $\omega_{\rm 1 eff}=\omega_1 +\frac{1}{3} \Gamma_2|A_{\rm ad}|^2$, coincides with $\omega_2/3$, in which case $\delta \Omega_{12} + |A_{\rm ad}|^2 = 0$. 

One can understand the mode-coupling induced relaxation in terms of the transitions into a thermal reservoir mimicked by mode 2. The energy relaxation rate of mode 1, which is $\propto  d|A_{\rm ad}|^2/dt$, has a term determined by the rate of such transitions. This rate is given by the Fermi golden rule. The effective density of states of the reservoir at the relevant frequency $3\omega_{\rm 1 eff}$ is $\propto \Gamma_2/[\Gamma_2^2 +(\omega_2 - 3\omega_{\rm 1 eff})^2]$. The transition rate is proportional to this density of states and is quadratic in the coupling $\propto \gamma_{12}^2(Q_1^2+P_1^2)^3$. The Fermi-golden-rule calculation is in full agreement with Eq.~(\ref{eq:adiabatic}). 

A nonexponential decay of the vibration amplitude is associated with a nonlinear friction force experienced by the vibrational system. Nonlinear friction is a topic of increasing interest in nano- and micro-mechanics, cf. \cite{Eichler2011a,Zaitsev2012,Imboden2013,Miao2014,Mahboob2015,polunin2016characterization,Atalaya2016b}. Usually it is described by a term  $\propto - |A_{\rm ad}|^3$  in the right-hand side of Eq.~(\ref{eq:adiabatic}). The fact that nonlinear coupling to a fast decaying mode can lead to such nonlinear friction was indicated, but not elaborated in Ref.~\cite{Dykman1975a}. In the present case the nonlinear friction is anomalously large and has a different dependence on the mode amplitude.

If the Duffing nonlinearity parameter $\gamma_{11}$ is small, for not too large vibration amplitudes one can disregard the term $\propto |A_{\rm ad}|^2$ in the denominator in Eq.~(\ref{eq:adiabatic}). The decay of the the scaled squared vibration amplitude $z = [\zeta_{\rm ad}/(1+\delta\Omega_{12}^2)]^{1/2} |A_{\rm ad}|^2$ is then described by a simple expression 
\begin{align*}
z(t) = z(0)\exp(-2\Gamma_1t)/\{1+z^2(0)[1-\exp(-4\Gamma_1 t)]\}^{-1/2}
\end{align*}
As expected, the decay is profoundly nonexponential in time. It is faster than exponential, and becomes exponential only for large time, where $z(t)\ll 1$.

\begin{figure}[h]
\includegraphics[width=2.5in]{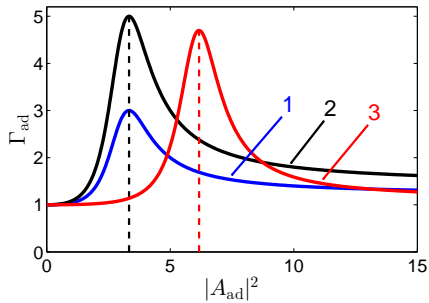}
\caption{The dependence of the effective instantaneous decay rate $\Gamma_{\rm ad}$ on the scaled vibration amplitude $|A_{\rm ad}|$ of mode 1  in the adiabatic regime of fast decaying mode 2. Curves 1 and 2 refer to   $\delta\Omega_{12} = -3$ and  $\zeta_{\rm ad} =0.2$ and 0.4, respectively. Curve 3 refers to $\delta\Omega_{12}=-6$ and $\zeta_{\rm ad} = 0.3$}
\label{fig:adiabatic_amplitude}
\end{figure}

For larger initial amplitudes or stronger Duffing nonlinearity the resonant nature of the nonlinear decay described by Eq.~(\ref{eq:adiabatic}) becomes more pronounced.  It is illustrated in Fig.~\ref{fig:adiabatic_amplitude} for the scaled instantaneous decay rate $\Gamma_{\rm ad} = \Gamma_1^{-1} d(\log |A_{\rm ad}|)/dt$. The figure shows two important features of the instantaneous decay rate. First, it displays a resonant peak where the instantaneous frequency of mode 1 $\omega_{1{\rm eff}}$ equals to $\omega_2/3$. The height of the peak increases with the increasing coupling strength $\zeta_{\rm ad}$ and with the increasing frequency detuning $|\delta\Omega_{12}| \propto \omega_2 - 3\omega_1$. Second, the instantaneous rate becomes constant not only for small vibration amplitudes, where $\Gamma_{\rm ad} \approx 1$, but also for $|A_{\rm ad}|^2\gg |\delta\Omega_{12}|$, where $\Gamma_{\rm ad} \approx 1+\zeta_{\rm ad}$. 

Interestingly, for relatively large amplitudes, where $|A_{\rm ad}|^2 +\delta\Omega_{12} \gg 1$, the detuning of the triple effective frequency of mode 1, $3\omega_{\rm 1 eff} - \omega_2$, largely exceeds the width  $\Gamma_2$  of the resonant response of mode 2. This means that  $3\omega_{\rm 1 eff}$ is on the tail of the Lorentzian ``density of states" of the effective thermal reservoir represented by mode 2. Yet, the decay rate $\Gamma_{\rm ad}$ does not fall off to the ``bare" decay rate $\Gamma_1$ in the absence of the mode coupling. This is a consequence of the strong nonlinearity of the  coupling, with the coupling energy $\propto |A_{\rm ad}|^3$.

\subsection{Effect of the strongly nonlinear frequency shift}
\label{subsec:dispersion}

Along with dissipation, the backaction from a thermal reservoir generally leads, in quantum terms, to a change of the energy spectrum. In classical terms, it corresponds to a change of the effective vibration frequency,  $\omega_{\rm 1 eff} \to \omega_{\rm 1 eff} + \Delta\omega_1$. This change in the adiabatic case is described by the imaginary part of ${\cal R}(|A_{\rm ad}|^2)$ in Eq.~(\ref{eq:adiabatic}),
\begin{align*}
 \Delta\omega_1 = -\Gamma_1\zeta_{\rm ad} |A_{\rm ad}|^4 {\rm Im}\{[1 + i(\delta\Omega_{12} +|A_{\rm ad}|^2)]^{-1}\}.
\end{align*}
The shift $\Delta\omega_1$  is a strongly nonlinear function of the amplitude, much stronger than the Duffing nonlinearity-induced shift, which is $\propto |A_{\rm ad}|^2$. In the interesting case $\delta\Omega_{12} <0$, the shift $\Delta\omega_1$ changes sign with the varying vibration amplitude. The sign change occurs at resonance, $3\omega_{\rm 1 eff}=\omega_2$. It is an analog of the level repulsion in quantum mechanics, which in this case looks like a turnaround of the vibration frequency.   It strongly affects the response of the mode 1 to the driving. 

Somewhat unexpectedly, for $\delta\Omega_{12} <0$ the slope of the frequency $\omega_{\rm 1 eff} + \Delta\omega_1$ as function of the amplitude can change twice with the increasing $|A_{\rm ad}|^2$. The first change, from the positive to the negative slope, can occur for $|A_{\rm ad}|^2\approx -\Gamma_2(1+\delta\Omega_{12}^2)/(6\Gamma_1\zeta_{\rm ad}\delta\Omega_{12})$ provided the corresponding $|A_{\rm ad}|^2$ is small compared to $-\delta\Omega_{12}$, i.e., for sufficiently strong initial detuning from the resonance (this is the most interesting case; the expression in the general case is elementary, but cumbersome). Then, with the increasing amplitude, the slope changes sign again before $|A_{\rm ad}|^2$ becomes larger than $|\delta\Omega_{12}|$. 

In the presence of driving, this unusual behavior leads to an unusual dependence of the amplitude $|A_{\rm ad}|$ on the parameters of the driving field. With the driving on, in the limit of fast decay of mode 2 the equation for $A_{\rm ad}$ reads
\begin{align}
\label{eq:adiabatic_drive}
\dot A_{\rm ad} =& -\Gamma_1 A_{\rm ad}{\cal R}(|A_{\rm ad}|^2)  -i\Gamma_2(3\beta_{\rm ad})^{1/2}
e^{-i\Phi_{\rm ad}(t)},
\end{align}
where $\beta_{\rm ad} = 3\gamma_{11}F^2/32\omega_F^3\Gamma_2^3$ is the scaled squared force amplitude; functions ${\cal R}$ and $\Phi_{\rm ad}$ are defined in Eqs.~(\ref{eq:amplitudes}) and (\ref{eq:adiabatic}).

Because of the nonlinear friction and the nonmonotonic dependence of the frequency on amplitude, the dynamics described by Eq.~(\ref{eq:adiabatic_drive}) is significantly more complicated than the dynamics of a Duffing oscillator. In dimensionless time $\Gamma_2 t$ it depends on five parameters. It is important that all parameters, except for the coupling constant $\gamma_{12}$, can be independently determined in the experiment by studying the dynamics of the modes for comparatively small drive, where the modes are essentially uncoupled.  

One of the first steps toward the understanding of the features of the dynamics related to the coupling is to study the stationary solution of Eq.~(\ref{eq:adiabatic_drive}). It determines how the amplitude and phase of forced vibrations vary with the varying amplitude $F$ and frequency $\omega_F$ of the drive. 

\begin{figure}[h]
\includegraphics[width=2.8in]{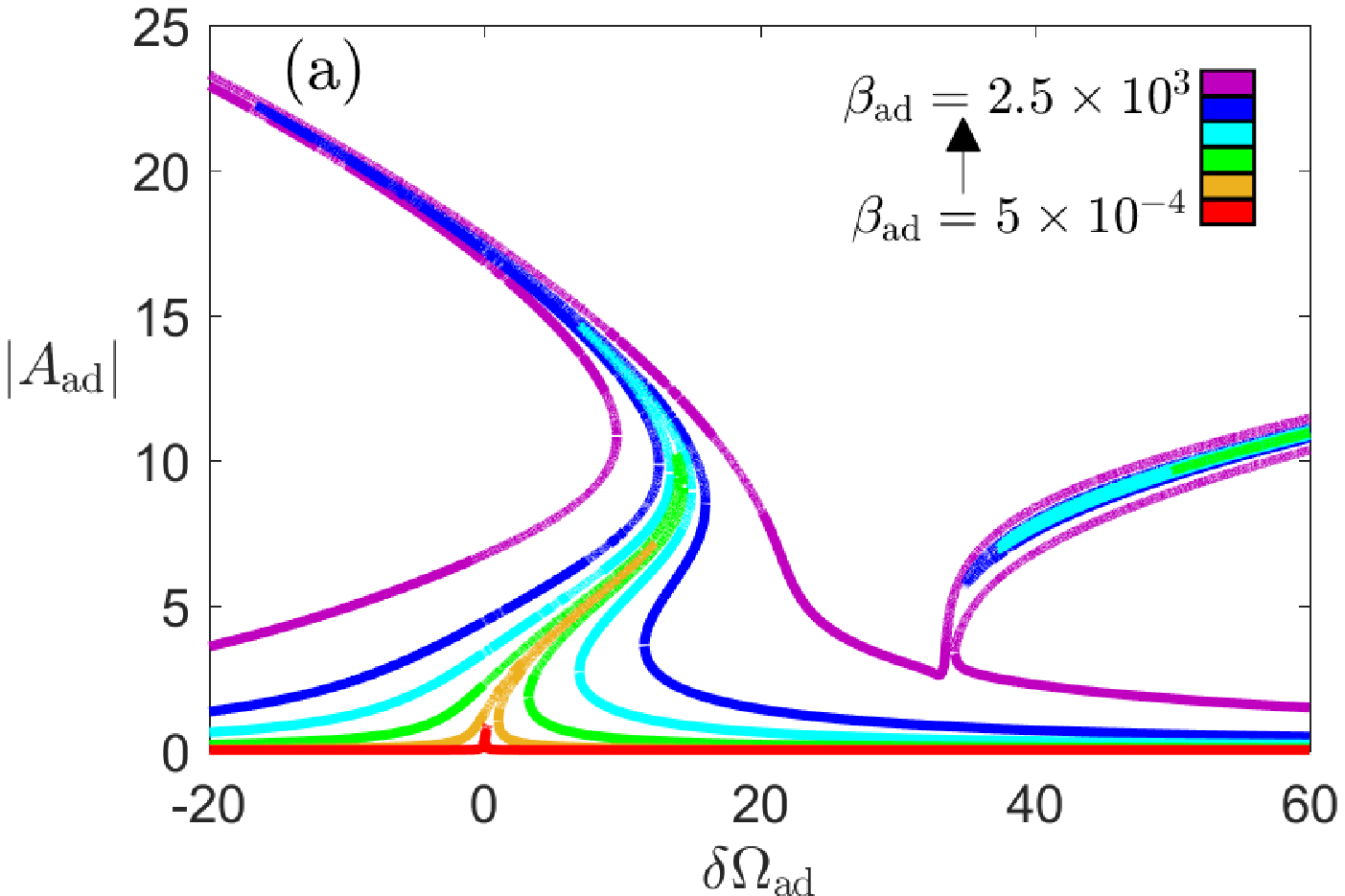}\\
\includegraphics[width=1.6in]{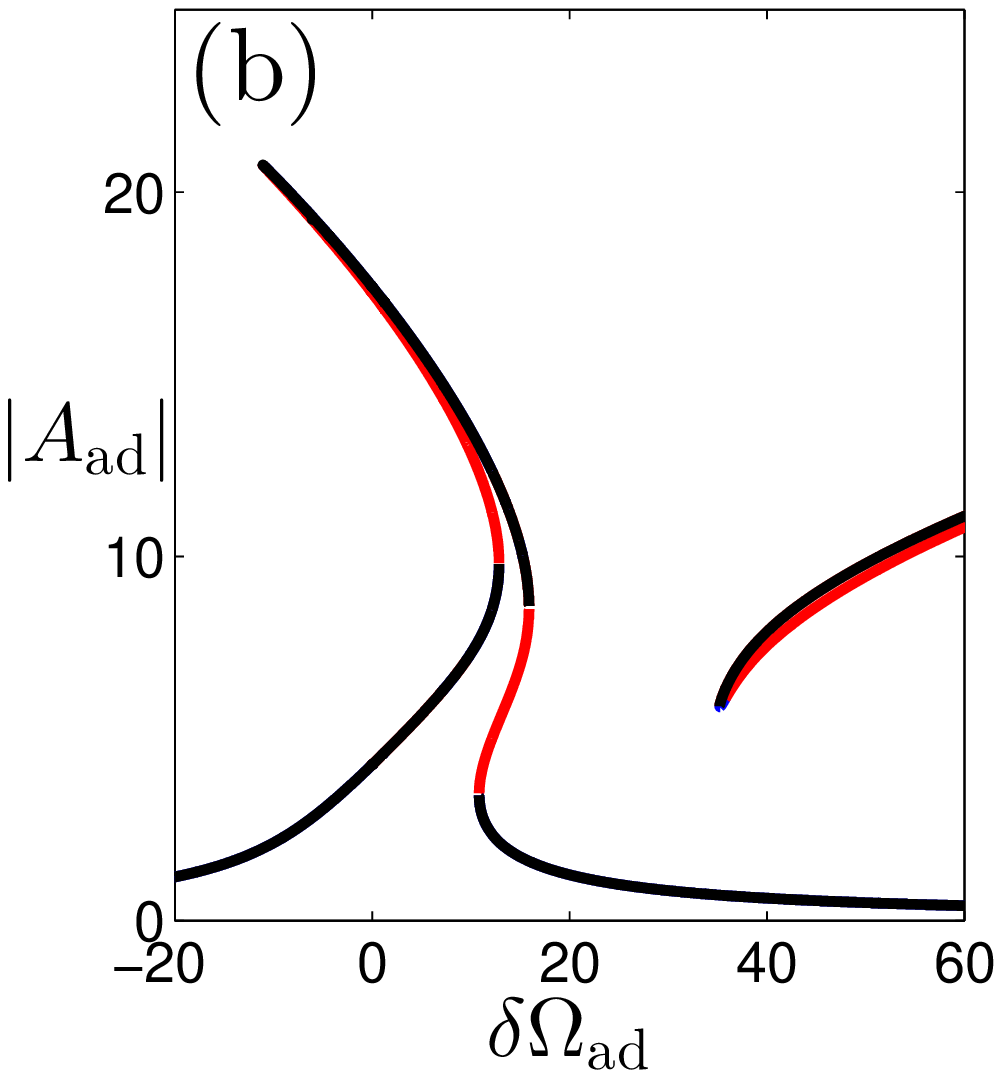}\hfill
\includegraphics[width=1.6in]{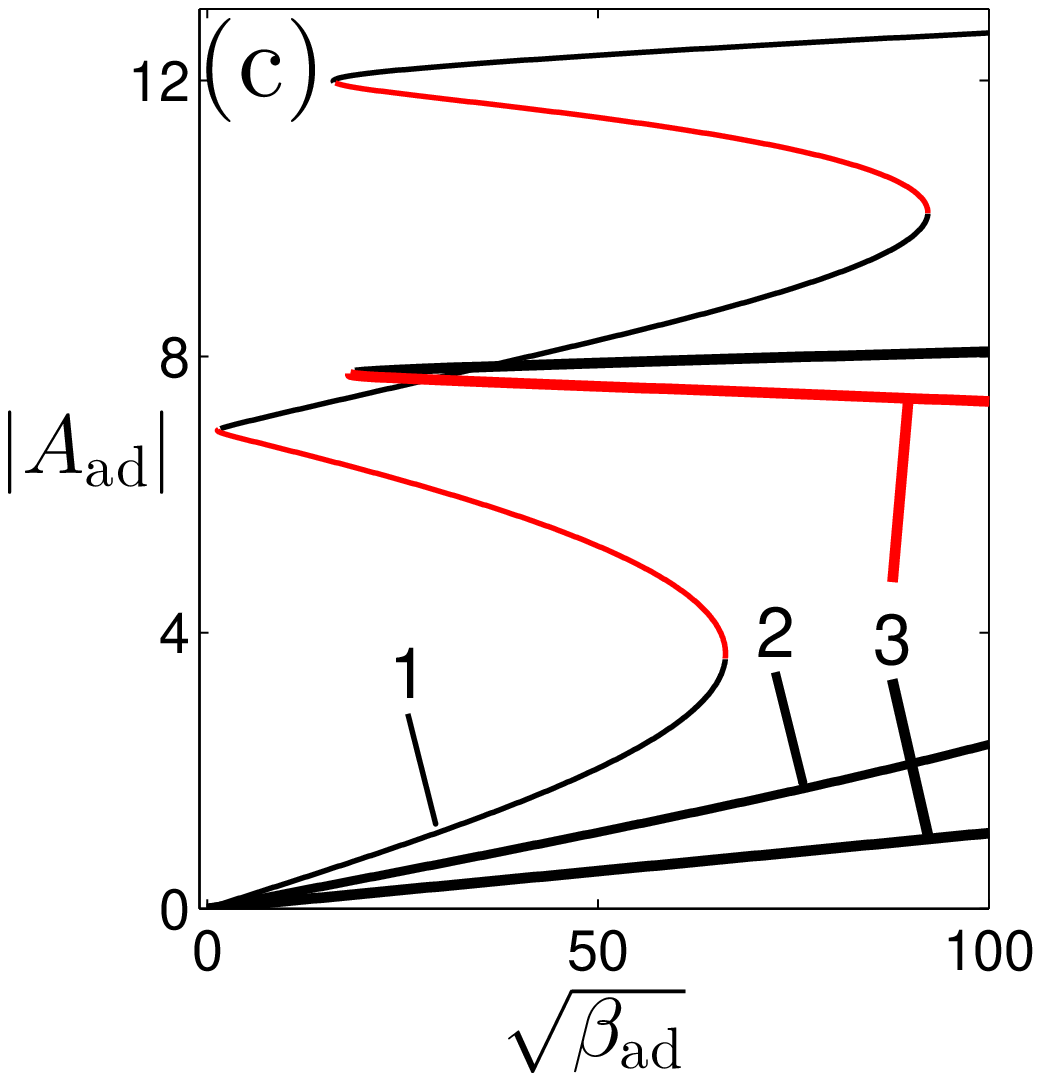}
\caption{(a) The amplitude of forced vibrations as a function of the scaled detuning of the drive frequency from the frequency of mode 1, $\delta\Omega_{\rm ad} = (\omega_F-\omega_1)/\Gamma_2$ and the scaled driving amplitude, $\beta_{\rm ad}=3\gamma_{11}F^2/32\omega_F^3\Gamma_2^3$. Other parameters are  $\gamma_{21}=3\gamma_{11}, (3\omega_{1}-\omega_2)/\Gamma_2=-100$, and $\Gamma_2/\Gamma_1 = 50$.  Function $|A_{\rm ad}|$ for particular parameter values is illustrated in (b) and (c). (b) Frequency dependence of the amplitude for $\beta_{\rm ad} = 320$. Black curves correspond to stable stationary states of forced vibrations, red curves show unstable stationary states. (c) the vibration amplitude as a function of the drive amplitude for a given frequency detuning $\delta\Omega_{\rm ad} = $~const. Curves 1, 2, and 3 refer to $\delta\Omega_{\rm ad}= 12, 20$ and 40. As in (b), black curves correspond to stable stationary states of forced vibrations, red curves show unstable stationary states. The results for extremely large $\beta_{\rm ad}$ illustrate the bifurcations in the system; they are of interest for the modes with extremely small decay rates $\Gamma_{1,2}$}
\label{fig:3D_amplitude}
\end{figure}

The amplitude of the forced vibrations is particularly important, as this is usually the first characteristic measured in the experiment. Equation (\ref{eq:adiabatic_drive}) has a form of the polynomial equation for the squared vibration amplitude of the stationary state $(|A_{\rm ad}|^2)_{\rm st} $,
\begin{align}
\label{eq:stationary}
x^2|{\cal R}(x)|^2 = 3\beta_{\rm ad}(\Gamma_2/\Gamma_1)^2, \quad (|A_{\rm ad}|^2)_{\rm st} = x.  
\end{align}
The solution of this equation as function of the scaled field amplitude $\beta_{\rm ad}$ and the scaled frequency detuning $\delta\Omega_{\rm ad}=(\omega_F-\omega_1)/\Gamma_2$ is shown in Fig.~\ref{fig:3D_amplitude}~(a). 

As seen from Fig.~\ref{fig:3D_amplitude}, the amplitude $(|A_{\rm ad}|)_{\rm st}$ has multiple branches and takes on different values for the same values of the drive parameters. The pattern becomes increasingly complicated as the system is brought into the nonlinear resonance by the increasing drive. This behavior is closely related to the nonmonotonous dependence of the ``vibration frequency'' $\omega_{\rm 1 eff} + \Delta\omega_1$ on the vibration amplitude discussed above and also to the strongly nonlinear friction. 

A part of the branches of  $(|A_{\rm ad}|)_{\rm st}$ correspond to stable states, whereas the other are unstable stationary states. Equation~(\ref{eq:adiabatic_drive}) describes the dynamics of a single complex variable or equivalently, two real variables. The stationary states of the system are either stable states or saddle points. Indeed, if we linearize Eq.~(\ref{eq:adiabatic_drive}) about a stationary state, we will see that the sum of the two eigenvalues of the corresponding characteristic equation is negative, which indicates that at least one of the eigenvalues that characterize the linearized motion has a negative real part. The system can switch between different stable states with varying parameters of the drive (or because of noise), but in the adiabatic limit there are no stationary states where the amplitude $A_{\rm ad}$ would oscillate. 

The stable and unstable branches of  $(|A_{\rm ad}|)_{\rm st}$ are shown in Fig.~\ref{fig:3D_amplitude} (b) and (c).  As seen from Fig.~\ref{fig:3D_amplitude} (b), as the driving frequency $\omega_F$ is varied, the amplitude can fold over, and there emerge regions of coexistence of 2 or 3 stable states of forced vibrations. These regions are separated by a gap where Eq.~(\ref{eq:adiabatic_drive}) has only one stationary solution. The gap is a consequence of the ``repulsion" of the vibration frequencies away from the resonance where $3\omega_{\rm 1 eff} = \omega_2$. It leads to an interesting hysteretic behavior. If we start from negative $\delta\Omega_{\rm ad}=(\omega_F-\omega_1)/\Gamma_2$ and increase $\omega_F$, we move from the small amplitude branch until it ends and we jump to the high-amplitude branch. From there, if we keep increasing $\omega_F$, we jump to the small-amplitude branch which goes to large positive $\delta\Omega_{\rm ad}$. The large amplitude branch for positive $\delta\Omega_{\rm ad}$ cannot be accessed by varying $\omega_F$ at all.

A complicated character of the hysteresis with varying field strength for constant $\omega_F$ is seen from Fig.~\ref{fig:3D_amplitude} (c).  On branch 1, which refers to the driving frequency in the region of the left ``beak" in Fig.~\ref{fig:3D_amplitude} (b), there are 3 coexisting stable states; the system switches with varying $F$ between the large/small amplitude branch and the intermediate-amplitude branch. Branch 2 refers to the gap in Fig.~\ref{fig:3D_amplitude} (b), where the system has only one branch. Branch 3 refers to the region of $\omega_F$ where there emerges the right ``beak" in Fig.~\ref{fig:3D_amplitude} (b) and the system has two branches of coexisting stable states. The merging of the unstable and low-amplitude branches occurs for very large $\beta_{\rm ad}$ and is not shown.

\section{Hamiltonian dynamics and the effect of weak dissipation}
\label{sec:Hamiltonian}

The dynamics of resonating modes is interesting also in the opposite limit where the decay of the modes can be disregarded all together. As seen from Eq.~(\ref{eq:RWA}), such dynamics is described by a time-independent Hamiltonian $H_{\rm RWA}$. It is convenient to re-write the equations of motion for scaled variables 
\begin{align}
\label{eq:scaled_variables}
&x_j = [3\gamma_{11}/8\omega_1^2\,(\delta\omega_1)]^{1/2}(Q_j - iP_j) \quad(j=1,2).
\end{align}
In dimensionless time $\tau = t\delta\omega_1$ in the neglect of the decay Eqs.~(\ref{eq:RWA}) read
\begin{align}
\label{eq:reduced_variables}
&dx_j/d\tau = 
- i\partial g/\partial x_j^*\quad (j=1,2)
\end{align}
with $g=g_0 + g_F$, 
\begin{align}
\label{eq:energy_reduced}
g_0=&|x_1|^2 +\frac{\delta\omega_2}{\delta\omega_1}|x_2|^2 -\frac{1}{2}|x_1|^4 -\frac{\gamma_{22}}{18\gamma_{11}}|x_2|^4 \nonumber\\
&-\alpha (x_1^3x_2^* + c.c), \qquad g_F=\beta^{1/2}(x_1+x_1^*). 
\end{align}
and
\begin{align}
\label{parameters}
\alpha = \gamma_{12}/3\sqrt{3}\gamma_{11}, \quad \beta = 3\gamma_11 F^2/32\omega_F^3(\delta\omega_1)^3
\end{align}

In the absence of the driving field $F\propto \beta^{1/2}$, Eqs.~(\ref{eq:energy_reduced}) provide a simple example of the standard nonlinear resonance. We reiterate that the analysis is grossly simplified by using the RWA. For $\beta=0$ the system has two integrals of motion, the effective  energy $g$ and the Manley-Rowe invariant $M=\frac{1}{3}|x_1|^2 + |x_2|^2 \propto Q_1^2+P_1^2 + 3(Q_2^2+P_2^2)$. We note that $g$ is not the energy of the oscillators. Rather it is an analog of the quasienergy of a periodically driven quantum system \cite{Zeldovich1967,Ritus1967,Sambe1973} in the limit of a vanishing field. 

The $\beta=0$-dynamics (\ref{eq:energy_reduced}) is thus integrable \cite{Arnold1989}. It maps on the dynamics of a pendulum. In the variables $I=\frac{1}{3}|x_1|^2$ and $\psi =  \arg x_2 -3\arg x_1$ the equations of motion read
\begin{align}
\label{eq:pendulum}
dI/d\tau = -\partial_\psi g_0(I,\psi), \quad  d\psi/d\tau = \partial_I g_0(I,\psi), 
\end{align}
where $g_0(I,\psi)$ is given by Eq.~(\ref{eq:energy_reduced}) in which $x_1$ and $ x_2$ are expressed in terms of $I,\psi, g_0$ and $M$.

Equations of motion (\ref{eq:pendulum}) can be explicitly solved in terms of the Jacobi elliptic functions. The dimensionless period of the vibrations  $T(g_0,M)$ can be expressed in terms of the complete elliptic integral of the first kind. On the phase plane $(I,\psi)$ the Hamiltonian trajectories (\ref{eq:pendulum}) have the familiar pattern of the sets of closed and open (along the $\psi$-axis) trajectories separated by  a separatrix loop. On this loop the period $T(g_0,M)$ diverges.

\subsection{Weak dissipation and weak driving}

\begin{figure}
\includegraphics[scale = 0.6]{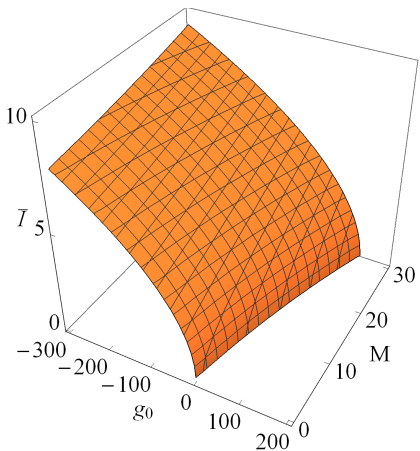}
\caption{The scaled squared amplitude $\overline{I}$ of vibrations of mode 1 averaged over the period $T(g_0,M)$ of vibrations in the rotating frame with given quasienergy $g_0$ and the Manley-Rowe invariant $M$. The data refer to $\gamma_{22}/\gamma_{11}=-1$ and $\alpha = 1/3$. }
\label{fig:I_boring}
\end{figure}

The Hamiltonian formulation allows us to describe the effect of weak dissipation and weak driving in a simple form. Away from the separatrix loop the dissipation is slow compared to the vibrations (\ref{eq:pendulum}) and the motion can be described in terms of the period-averaged dynamical variables. We denote period-averaged variables by the overline. Of special interest to us is the period-averaged squared amplitude of mode 1, which is given by the scaled variable 
\begin{align}
\label{eq:averaging_defined}
\overline{I} \equiv \overline {I(g_0,k)} =T^{-1}(g_0,M)\int_0^{T(g_0,M)} d\tau  I(\tau|g_0,M)
\end{align}
with $I(\tau|g_0,M)$ given by the solution of Eq.~(\ref{eq:pendulum}) for given $g_0,M$. An example of this function is shown in Fig.~\ref{fig:I_boring}.

The evolution of $\overline{I}$ due to the dissipation is controlled by the evolution of the quasienergy $g_0$ and the invariant $M$. The latter is described by equation
\begin{align}
\label{eq:evolution}
\frac{dg_0}{d\tau} \equiv \left(\frac{dg_0}{d\tau}\right)_d= -(\delta\omega_1)^{-1}\sum_j \Gamma_j \overline{x_j\partial_{x_j} g_0}
+ {\rm c.c.}
\end{align}
and a similar equation for $M$. One can easily see that the right-hand sides of Eq.~(\ref{eq:evolution}) and the similar equation for $M$ are functions of $g_0$ and $M$. Once they are found, one can find the time evolution of the scaled squared vibration amplitude $\overline{I}$ from Eq.~(\ref{eq:averaging_defined}). This is a theoretical basis for describing ring-down measurements for weakly damped modes that undergo nonlinear resonance. The examples will be presented in a separate paper.  

In the presence of driving, the Hamiltonian system with quasienergy $g_0+g_F$ is no longer integrable. The analysis is simplified in the case of weak driving and weak dissipation. The driving pumps energy into the system, and the dissipation provides an energy drain. They compensate each other in the stationary regime. When both are slow compared to the vibration period $T(g_0,M)$, the system reaches a regime of periodic vibrations with the values of $g_0,M$ determined by the balance equation $(dg_0/d\tau)_d + (dg_0/d\tau)_F=0$, and a similar equation for $M$. In contrast to the dissipative term, the driving-induced change of the quasienergy $ (dg_0/d\tau)_F$ has to be calculated to the second order in the $\beta^{1/2}$ and then averaged over the period. 

An important immediate consequence of the presented analysis is that, for weak damping, the stationary regime of the coupled modes corresponds to vibrations of the mode amplitudes in the rotating frame. These vibrations have the period $T_0(g_0,M)$, which is much larger than the driving period $2\pi/\omega_F$.  They correspond to sustained amplitude modulated responses in the driven system, which often appear via Hopf bifurcations and can ultimately evolve into chaos \cite{bajaj1990torus,Sun2016,Mahboob2016}.

\section{Conclusions}
The presented analysis describes the dynamics of systems with weakly damped vibrational modes that experience nonlinear resonance. The specific results refer to 1:3 resonance. The peculiar features of the vibrational dynamics come from the strong dependence of the coupling strength on the mode amplitude and the amplitude dependence of the mode frequencies, which leads to tuning the modes into and out of  resonance as their amplitudes vary. Another factor is the interplay of the nonlinearity and the decay. 

The nonlinear resonant intermodal coupling results in non-exponential decay of the vibrations, with the decay rate that strongly depends on amplitude. It also leads to a non-monontonic dependence of the effective vibration frequency on amplitude even where, in the absence of the modal coupling,  this dependence is monotonic and comes from the simple Duffing nonlinearity.    The  decay rate and the frequency dependence on amplitude could be obtained  in an explicit form in the important case where the high-frequency mode decays much faster than the low-frequency one, and thus the high-frequency mode adiabatically follows the low-frequency mode. Resonant driving in this case leads to coexistence of several stable states of forced vibrations. The switching between the branches of the stable states with varying parameters of the drive has a peculiar pattern, where some branches can be reached by varying only the drive strength, but not the frequency, for example. The analysis indicates that the stable states correspond to vibrations with time-independent amplitudes.

In contrast, if the decay rates of the modes are small compared to the nonlinearity, in the absence of the drive the dynamics in the rotating frame  can be mapped onto a pendulum. With weak dissipation and weak drive, the stable states correspond to nonsinusoidal oscillations of the amplitude. The decay of the vibration amplitude after the drive is switched off is also profoundly nonexponential and can be accompanied by oscillations of the amplitude. It is described in terms of the decay of the effective quasi-energy and the Manley-Rowe invariant.

We are grateful to Adrian Bachtold, Changyao Chen, Dave Czaplewski,  Andreas Isacsson,   Daniel López, Pavel Polunin, and Scott Strachan for fruitful discussions. This work was supported in part by the US Army Research Office (W911NF-12-1-0235) and the US Defense Advanced Research Projects Agency (FA8650-13-1-7301).  S.W.S acknowledges partial support from the Natioinal Science Foundation (Grant No.CMMI-SDC 1561934).  O.S. and S.W.S were supported in part by Florida Institute of Technology. M. I. D. acknowledges partial support from the  National Science Foundation (Grant No. DMR-1514591).)

\bibliographystyle{apsrev4-1}
%

\end{document}